\documentclass[a4paper,12pt]{article}
\usepackage{amssymb}
\usepackage[latin1]{inputenc}     

\newcommand{\beq}{\begin{equation} }
\newcommand{\eqq}{\end{equation} }
\newcommand{\cuad}{{\sqcap\kern-.68em\sqcup}}

\newtheorem{remark}{Remark}[section]
\newcommand{\bremark}{\begin{remark} \em}
\newcommand{\eremark}{\end{remark} }

\def\beeq{\begin{equation}}
\def\eeq{\end{equation}}
\newcommand{\begeqaet}{\begin{eqnarray*}}
\newcommand{\eneqaet}{\end{eqnarray*}}

\let\Section=\section
\def\section{\setcounter{equation}{0}\Section}
\newtheorem{Lem}{Lemma}[section]
\newtheorem{Thm}{Theorem}[section]
\newtheorem{Def}{Definition}[section]

\newtheorem{Remark}{Remark}[section]

\begin{document}
\begin{center}{\bf\Large Multiplicity of solutions for fractional Hamiltonian systems with Liouville-Weyl fractional derivative}\medskip

\bigskip

\bigskip

{Amado M\'endez and C\'esar Torres}

 Departamento Acad\'emico de Matem\'aticas\\ 
 Universidad Nacional de Trujillo\\
Av. Juan Pablo s/n, Trujillo, Per\'u.\\
 {\sl  (gamc55@hotmail.com, ctl\_576@yahoo.es)}

\end{center}

\medskip

\medskip
\medskip
\medskip
\medskip

\begin{abstract}
In this paper, we investigate the existence of infinitely many solutions for the following fractional Hamiltonian systems:
\begin{eqnarray}\label{eq00}
_{t}D_{\infty}^{\alpha}(_{-\infty}D_{t}^{\alpha}u(t)) + L(t)u(t)  = & \nabla W(t,u(t))\\
u\in H^{\alpha}(\mathbb{R}, \mathbb{R}^{N}).\nonumber
\end{eqnarray}
where $\alpha \in (1/2, 1)$, $t\in \mathbb{R}$, $u\in \mathbb{R}^{n}$, $L\in C(\mathbb{R}, \mathbb{R}^{n^2})$ is a symmetric and positive definite matrix for all $t\in \mathbb{R}$, $W\in C^{1}(\mathbb{R}\times \mathbb{R}^{n}, \mathbb{R})$, and $\nabla W$ is the gradient of $W$ at $u$. The novelty of this paper is that, assuming there exists $l\in C(\mathbb{R}, \mathbb{R})$ such that $(L(t)u,u)\geq l(t)|u|^{2}$ for all $t\in \mathbb{R}$, $u\in \mathbb{R}^{n}$ and the following conditions on $l$: $\inf_{t\in \mathbb{R}}l(t) >0$ and there exists $r_{0}>0$ such that, for any $M>0$
    $$
    m(\{t\in (y-r_{0}, y+r_{0})/\;\;l(t)\leq M\}) \to 0\;\;\mbox{as}\;\;|y|\to \infty.
    $$
   are satisfied and $W$ is of subquadratic growth as $|u| \to +\infty$, we show that (\ref{eq00}) possesses infinitely many solutions via the genus properties in the critical theory. Recent results in [Z. Zhang and R. Yuan, Solutions for subquadratic fractional Hamiltonian systems without coercive conditions, Math. Methods Appl. Sci., DOI: 10.1002/mma.3031] are significantly improved.
  
\noindent
{\bf Key works:} Liouville-Weyl fractional derivative, fractional Hamiltonian systems, critical point, variational methods.
\end{abstract}
\date{}

\setcounter{equation}{0}
\section{ Introduction}

Fractional differential equations both ordinary and partial ones are applied in mathematical modeling of processes in
physics, mechanics, control theory, biochemistry, bioengineering and economics. Therefore the theory of fractional
differential equations is an area intensively developed during last decades \cite{OAJTMJS}, \cite{RH}, \cite{AKHSJT}, \cite{RMJK}, \cite{KMBR}, \cite{IP}, \cite{JSOAJTM}, \cite{BWMBPG}. Therein, the composition of fractional differential operators has got much attention from many scientists, mainly due to its wide applications in modeling physical phenomena exhibiting anomalous diffusion. Specifically, the models involving a fractional differential oscillator equation, which contains a composition of left and right fractional derivatives, are proposed for the description of the processes of emptying the silo \cite{JLTB} and the heat flow through a bulkhead filled with granular material \cite{ES}, respectively. Their studies show that the proposed models based on fractional calculus are efficient and describe well the processes.

In the aspect of theory, the study of fractional differential equations including both left and right fractional derivatives has attracted much attention by using fixed point theory and variational methods \cite{TABS}, \cite{DBJT}, \cite{MK}, \cite{CT}, \cite{CT1}, \cite{CT2}, \cite{CT3}, \cite{CT4}, \cite{ZZRY} and their references. We note that, it is not easy to use the critical point theory to study the fractional differential equations including both left and right fractional derivatives, since it is often very difficult to establish a suitable space and a variational functional for the fractional boundary value problem.

Very recently in \cite{CT} the author considered the following fractional Hamiltonian systems
\begin{eqnarray}\label{Eq02}
_{t}D_{\infty}^{\alpha}(_{-\infty}D_{t}^{\alpha}u(t)) + L(t)u(t)  = & \nabla W(t,u(t))
\end{eqnarray}
where $\alpha \in (1/2,1)$, $t\in \mathbb{R}$, $u\in \mathbb{R}^{n}$, $L\in C(\mathbb{R}, \mathbb{R}^{n^2})$ is a symmetric matrix valued function for all $t\in \mathbb{R}$, $W\in C^{1}(\mathbb{R}\times \mathbb{R}^{n}, \mathbb{R})$ and $\nabla W(t, u(t))$ is the gradient of $W$ at $u$. Assuming that $L$ and $W$ satisfy the following hypotheses: 
\begin{itemize}
\item[$(L)$] $L(t)$ is positive definite symmetric matrix for all $t\in \mathbb{R}$, and there exists an $l\in C(\mathbb{R}, (0,\infty))$ such that $l(t) \to +\infty$ as $t \to \infty$ and
    \begin{equation}\label{Eq03}
    (L(t)x,x) \geq l(t)|x|^{2},\;\;\mbox{for all}\;t\in \mathbb{R}\;\;\mbox{and}\;\;x\in \mathbb{R}^{n}.
    \end{equation}
\item[$(W_{1})$] $W\in C^{1}(\mathbb{R} \times \mathbb{R}^{n}, \mathbb{R})$, and there is a constant $\mu >2$ such that
$$
0< \mu W(t,x) \leq (x, \nabla W(t,x)),\;\;\mbox{for all}\;t\in \mathbb{R}\;\;\mbox{and}\;x\in\mathbb{R}^{n}\setminus \{0\}.
$$
\item[$(W_{2})$] $|\nabla W(t,x)| = o(|x|)$ as $x\to 0$ uniformly with respect to $t\in \mathbb{R}$.
\item[$(W_{3})$] There exists $\overline{W} \in C(\mathbb{R}^{n}, \mathbb{R})$ such that
$$
|W(t,x)| + |\nabla W(t,x)| \leq |\overline{W(x)}|\;\;\mbox{for every}\;x\in \mathbb{R}^{n}\;\mbox{and}\;t\in \mathbb{R}.
$$
\end{itemize}
It showed that (\ref{Eq02}) has at least one nontrivial solution via Mountain pass theorem. 

In particular, if $\alpha = 1$, (\ref{Eq02}) reduces to the standard second order differential equation
\begin{equation}\label{HEq01}
u'' - L(t)u + \nabla W(t,u)=0,
\end{equation}
where $W: \mathbb{R} \times \mathbb{R}^{n} \to \mathbb{R}$ is a given function and $\nabla W(t,u)$ is the gradient of $W$ at $u$. The existence of homoclinic solution is one of the most important problems in the history of that kind of equations, and has been studied intensively by many mathematicians. Assuming that $L(t)$ and $W(t,u)$ are independent of $t$, or $T$-periodic in $t$, many authors have studied the existence of homoclinic solutions for (\ref{HEq01}) via critical point theory and variational methods. In this case, the existence of homoclinic solution can be obtained by going to the limit of periodic solutions of approximating problems.

If $L(t)$ and $W(t,u)$ are neither autonomous nor periodic in $t$, this problem is quite different from the ones just described, because the lack of compacteness of the Sobolev embedding. In \cite{PRKT} the authors considered (\ref{HEq01}) without periodicity assumptions on $L$ and $W$ and showed that (\ref{HEq01}) possesses one homoclinic solution by using a variant of the mountain pass theorem without the Palais-Smale contidion. In \cite{WOMW}, under the same assumptions of \cite{PRKT}, the authors, by employing a new compact embedding theorem, obtained the existence of homoclinic solution of (\ref{HEq01}).

Motivated by the previously mentioned results, using the genus properties of critical point theory, in \cite{ZZRY}, the authors generalized the result of \cite{CT} and established some new criterion to guarantee the existence of infinitely many solutions of (\ref{Eq02}) for the case that $W(t,u)$ is subquadratic as $|u| \to +\infty$. Explicitly, $L$ satisfies $(L)$ and the potential $W(t,u)$ is supposed to satisfy the following conditions:
\begin{itemize}
\item[$(HS)_{1}$] $W(t,0) = 0$ for all $t\in \mathbb{R}$, $W(t,u) \geq a(t)|u|^{\theta}$ and $|\delta W(t,u)| \leq b(t)|u|^{\theta -1}$ for all $(t,u)\in \mathbb{R} \times \mathbb{R}^{n}$, where $ \theta < 2$ is a constant, $a :\mathbb{R} \to \mathbb{R}^{+}$ is a bounded continuous function and $b:\mathbb{R} \to \mathbb{R}^{+}$ is a continuous function such that $b\in L^{\frac{2}{2-\theta}}(\mathbb{R})$;
\item[$(HS)_{2}$] There is a constant $1 < \sigma \leq \theta < 2$ such that
$$
(W(t,u), u) \leq \sigma W(t,u)\quad \mbox{for all t}\in \mathbb{R}\;\;\mbox{and}\;\;u\in \mathbb{R}^{n}\setminus \{0\};
$$
\item[$(HS)_{3}$] $W(t,u)$ is even in $u$, i.e. $W(t,u) = W(t,-u)$ for all $t\in \mathbb{R}$ and $u\in \mathbb{R}^{n}$.
\end{itemize}
Moreover they got the behavior
\begin{equation}\label{Beq04}
\int_{\mathbb{R}} \left[ \frac{1}{2}|_{-\infty}D_{t}^{\alpha}u_{j}(t)|^2 + \frac{1}{2} (L(t)u_{j}(t), u_{j}(t)) - W(t,u_{j}(t))\right]dt \to 0^{-}
\end{equation}
as $j\to +\infty$.

As is well-known, the condition $(L)$ is the so-called coercive condition and is a little demanding. In fact, for a simple choice like $L(t) = sId_{n}$, the condition (\ref{Eq03}) is not satisfied, where $s>0$ and $Id_{n}$ is the $n\times n$ identity matrix. Considering this trouble, very recently in \cite{ZZRY1}  the recent results in \cite{ZZRY} are generalized and significantly improved. More precisely  in \cite{ZZRY1} the authors considered the case that L(t) is bounded in the sense that
\begin{itemize}
\item[$(L)'$] $L\in C(\mathbb{R}, \mathbb{R}^{n^2})$ is a symmetric and positive definite matrix for all $t\in \mathbb{R}$ and there are constants $0 < \tau_{1} < \tau_{2} <+\infty$ such that
$$
\tau_{1}|u|^2 \leq (L(t)u,u) \leq \tau_{2}|u|^2\;\;\mbox{for all} \;\;(t,u)\in \mathbb{R}\times \mathbb{R}^n.
$$
\end{itemize}
and the potential $W(t,u)$ is supposed to satisfy the following assumptions:
\begin{itemize}
\item[$(HS)'_{1}$] $W(t,0) = 0$ for all $t\in \mathbb{R}$, $W(t,u) \geq a(t)|u|^{\theta}$ and $|\nabla W(t,u)| \leq b(t)|u|^{\theta - 1}$ for all $(t,u)\in \mathbb{R} \times \mathbb{R}^n$, where $1<\theta <2$ is a constant, $a:\mathbb{R} \to \mathbb{R}^{+}$ is a bounded continuous function, and $b:\mathbb{R} \to \mathbb{R}^{+}$ is a continuous function such that $b\in L^{\xi}(\mathbb{R}, \mathbb{R})$ for some $1\leq \xi \leq 2$.
\end{itemize}

\noindent 
Under conditions $(L)', (HS)'_{1}$ and $(HS)_{3}$, the authors proved that (\ref{Eq02}) has infinitely many nontrivial solutions. But they lost the behavior (\ref{Beq04}).

Motivated by this previous result, in this paper we consider the existence of infinitely many nontrivial solution to (\ref{Eq02}) under some weaker condition than $(L)$ and we recovered the behavior (\ref{Beq04}). More precisely we consider
\begin{itemize}
\item[$(L_{w})$] $L(t)$ is positive definite symmetric matrix for all $t\in \mathbb{R}$, and there exists an $l\in C(\mathbb{R}, \mathbb{R})$ such that
 \begin{itemize}
    \item[$(L_{w}^{1})$] $\inf_{t\in \mathbb{R}}l(t) >0$,
    \item[$(L_{w}^{2})$] There exists $r_{0}>0$ such that, for any $M>0$
    $$
    m(\{t\in (y-r_{0}, y+r_{0})/\;\;l(t)\leq M\}) \to 0\;\;\mbox{as}\;\;|y|\to \infty.
    $$
    \end{itemize}
   and 
    \begin{equation}\label{Eq03}
    (L(t)x,x) \geq l(t)|x|^{2},\;\;\mbox{for all}\;t\in \mathbb{R}\;\;\mbox{and}\;\;x\in \mathbb{R}^{n}.
    \end{equation}
\end{itemize}     

Up until now, we can state our main resut
\begin{Thm}\label{tm01}
Suppose that $(L_{w})$, $(HS)_{1}-(HS)_{3}$ are satisfied. Then, (\ref{Eq02}) has infinitely many nontrivial solutions $\{u_{j}\}_{j\in \mathbb{N}}$ such that
$$
\int_{\mathbb{R}} \left[ \frac{1}{2}|_{-\infty}D_{t}^{\alpha}u_{j}(t)|^2 + \frac{1}{2} (L(t)u_{j}(t), u_{j}(t)) - W(t,u_{j}(t))\right]dt \to 0^{-}
$$ 
as $j\to +\infty$.
\end{Thm}

\begin{Remark}\label{nta01}
From $(HS)_{1}$, it is easy to check that $W(t,u)$ is subquadratic as $|u| \to +\infty$. In fact, in view of $(HS)_{1}$, we have  
\begin{equation}\label{Eq04}
W(t,u) = \int_{0}^{1}(\nabla W(t,su), u)ds \leq \frac{b(t)}{\theta} |u|^{\theta},
\end{equation}
which implies that $W(t,u)$ is of subquadratic growth as $|u|\to +\infty$. 

\end{Remark}
\begin{Remark}\label{nta02}
In \cite{CT}, assuming $(L)$ holds, the author introduced some compact embedding lemma (see its Lemma 2.2), which has also been used in \cite{ZZRY} to verify that the corresponding functional of (\ref{Eq02}) satisfies the (PS) condition. In our present paper, we weaken $(L)$ to $(L_{w})$ and under this new condition we get a new compact embedding result (see Lemma \ref{FDElm02}). 
\end{Remark}

The rest of the paper is organized as follows: Some preliminary results are presented in Section \S 2. In Section \S 3, we are devoted to accomplishing the proof of our main result.

\section{Preliminary Results}

\subsection{Liouville-Weyl Fractional Calculus}
In this section we introduce some basic definitions of fractional calculus which are used further in this paper. For more details we refer the reader to \cite{RH}. 

The Liouville-Weyl fractional integrals of order $0<\alpha < 1$ are defined as
\begin{equation}\label{LWeq01}
_{-\infty}I_{x}^{\alpha}u(x) = \frac{1}{\Gamma (\alpha)} \int_{-\infty}^{x}(x-\xi)^{\alpha - 1}u(\xi)d\xi
\end{equation}
\begin{equation}\label{LWeq02}
_{x}I_{\infty}^{\alpha}u(x) = \frac{1}{\Gamma (\alpha)} \int_{x}^{\infty}(\xi - x)^{\alpha - 1}u(\xi)d\xi
\end{equation}
The Liouville-Weyl fractional derivative of order $0<\alpha <1$ are defined as the left-inverse operators of the corresponding Liouville-Weyl fractional integrals
\begin{equation}\label{LWeq03}
_{-\infty}D_{x}^{\alpha}u(x) = \frac{d }{d x} {_{-\infty}}I_{x}^{1-\alpha}u(x)
\end{equation}
\begin{equation}\label{LWeq04}
_{x}D_{\infty}^{\alpha}u(x) = -\frac{d }{d x} {_{x}}I_{\infty}^{1-\alpha}u(x)
\end{equation}
The definitions (\ref{LWeq03}) and (\ref{LWeq04}) may be written in an alternative form:
\begin{equation}\label{LWeq05}
_{-\infty}D_{x}^{\alpha}u(x) = \frac{\alpha}{\Gamma (1-\alpha)} \int_{0}^{\infty}\frac{u(x) - u(x-\xi)}{\xi^{\alpha + 1}}d\xi
\end{equation}
\begin{equation}\label{LWeq05}
_{x}D_{\infty}^{\alpha}u(x) = \frac{\alpha}{\Gamma (1-\alpha)} \int_{0}^{\infty}\frac{u(x) - u(x+\xi)}{\xi^{\alpha + 1}}d\xi
\end{equation}

\noindent
We establish the Fourier transform properties of the fractional integral and fractional differential operators. Recall that the Fourier transform $\widehat{u}(w)$ of $u(x)$ is defined by
$$
\widehat{u}(w) = \int_{-\infty}^{\infty} e^{-ix.w}u(x)dx.
$$
Let $u(x)$ be defined on $(-\infty, \infty)$. Then the Fourier transform of the Liouville-Weyl integral and differential operator satisfies
\begin{equation}\label{LWeq06}
\widehat{ _{-\infty}I_{x}^{\alpha}u(x)}(w) = (iw)^{-\alpha}\widehat{u}(w)
\end{equation}
\begin{equation}\label{LWeq07}
\widehat{ _{x}I_{\infty}^{\alpha}u(x)}(w) = (-iw)^{-\alpha}\widehat{u}(w)
\end{equation}
\begin{equation}\label{LWeq08}
\widehat{ _{-\infty}D_{x}^{\alpha}u(x)}(w) = (iw)^{\alpha}\widehat{u}(w)
\end{equation}
\begin{equation}\label{LWeq09}
\widehat{ _{x}D_{\infty}^{\alpha}u(x)}(w) = (-iw)^{\alpha}\widehat{u}(w)
\end{equation}
\subsection{Fractional Derivative Spaces}

In this section we introduce some fractional spaces for more detail see \cite{VEJR}.

\noindent
Let $\alpha > 0$. Define the semi-norm
$$
|u|_{I_{-\infty}^{\alpha}} = \|_{-\infty}D_{x}^{\alpha}u\|_{L^{2}}
$$
and norm
\begin{equation}\label{FDEeq01}
\|u\|_{I_{-\infty}^{\alpha}} = \left( \|u\|_{L^{2}}^{2} + |u|_{I_{-\infty}^{\alpha}}^{2} \right)^{1/2},
\end{equation}
and let
$$
I_{-\infty}^{\alpha} (\mathbb{R}) = \overline{C_{0}^{\infty}(\mathbb{R})}^{\|.\|_{I_{-\infty}^{\alpha}}}.
$$
Now we define the fractional Sobolev space $H^{\alpha}(\mathbb{R})$ in terms of the fourier transform. Let $0< \alpha < 1$, let the semi-norm
\begin{equation}\label{FDEeq02}
|u|_{\alpha} = \||w|^{\alpha}\widehat{u}\|_{L^{2}}
\end{equation}
and norm
$$
\|u\|_{\alpha} = \left( \|u\|_{L^{2}}^{2} + |u|_{\alpha}^{2} \right)^{1/2},
$$
and let
$$
H^{\alpha}(\mathbb{R}) = \overline{C_{0}^{\infty}(\mathbb{R})}^{\|.\|_{\alpha}}.
$$

\noindent
We note a function $u\in L^{2}(\mathbb{R})$ belong to $I_{-\infty}^{\alpha}(\mathbb{R})$ if and only if
\begin{equation}\label{FDEeq03}
|w|^{\alpha}\widehat{u} \in L^{2}(\mathbb{R}).
\end{equation}
Especially
\begin{equation}\label{FDEeq04}
|u|_{I_{-\infty}^{\alpha}} = \||w|^{\alpha}\widehat{u}\|_{L^{2}}.
\end{equation}
Therefore $I_{-\infty}^{\alpha}(\mathbb{R})$ and $H^{\alpha}(\mathbb{R})$ are equivalent with equivalent semi-norm and norm. Analogous to $I_{-\infty}^{\alpha}(\mathbb{R})$ we introduce $I_{\infty}^{\alpha}(\mathbb{R})$. Let the semi-norm
$$
|u|_{I_{\infty}^{\alpha}} = \|_{x}D_{\infty}^{\alpha}u\|_{L^{2}}
$$
and norm
\begin{equation}\label{FDEeq05}
\|u\|_{I_{\infty}^{\alpha}} = \left( \|u\|_{L^{2}}^{2} + |u|_{I_{\infty}^{\alpha}}^{2} \right)^{1/2},
\end{equation}
and let
$$
I_{\infty}^{\alpha}(\mathbb{R}) = \overline{C_{0}^{\infty}(\mathbb{R})}^{\|.\|_{I_{\infty}^{\alpha}}}.
$$
Moreover $I_{-\infty}^{\alpha}(\mathbb{R})$ and $I_{\infty}^{\alpha}(\mathbb{R})$ are equivalent , with equivalent semi-norm and norm \cite{VEJR}.

Now we recall the Sobolev lemma.
\begin{Thm}\label{FDEtm01}
\cite{CT} If $\alpha > \frac{1}{2}$, then $H^{\alpha}(\mathbb{R}) \subset C(\mathbb{R})$ and there is a constant $C=C_{\alpha}$ such that
\begin{equation}\label{FDEeq06}
\sup_{x\in \mathbb{R}} |u(x)| \leq C \|u\|_{\alpha}
\end{equation}
\end{Thm}

\begin{Remark}\label{FDEnta01}
From (\ref{FDEtm01}), we now that if $u\in H^{\alpha}(\mathbb{R})$ with $1/2 < \alpha <1$, then $u\in L^{q}(\mathbb{R})$ for all $q\in [2,\infty)$, because
$$
\int_{\mathbb{R}} |u(x)|^{q}dx \leq \|u\|_{\infty}^{q-2}\|u\|_{L^{2}}^{2}.
$$
\end{Remark}

\noindent
In what follows, we introduce the fractional space in which we will construct the variational framework of (\ref{Eq02}). Let
$$
X^{\alpha} = \left\{ u\in H^{\alpha}(\mathbb{R}, \mathbb{R}^{n})|\;\;\int_{\mathbb{R}} \left[|_{-\infty}D_{t}^{\alpha}u(t)|^{2} + (L(t)u(t),u(t))\right] dt < \infty  \right\},
$$
then $X^{\alpha}$ is a reflexive and separable Hilbert space with the inner product
$$
\langle u,v \rangle_{X^{\alpha}} = \int_{\mathbb{R}} \left[ (_{-\infty}D_{t}^{\alpha}u(t) , \; _{-\infty}D_{t}^{\alpha}v(t)) + (L(t)u(t),v(t))\right]dt
$$
and the corresponding norm
$$
\|u\|_{X^{\alpha}}^{2} = \langle u,u \rangle_{X^{\alpha}}
$$
Similar to Lemma 2.1 in \cite{CT}, we have the following conclusion. Its proof is just the repetition of Lemma 2.1 of \cite{CT}, so we omit the details.
\begin{Lem}\label{FDElm01}
Suppose $L$ satisfies ($L_{w}$). Then $X^{\alpha}$ is continuously embedded in $H^{\alpha}(\mathbb{R},\mathbb{R}^{n})$.
\end{Lem}

\begin{Lem}\label{FDElm02}
Suppose $L$ satisfies ($L_{w}$). Then the imbedding of $X^{\alpha}$ in $L^{2}(\mathbb{R}, \mathbb{R}^n)$ is compact.
\end{Lem}

\noindent
{\bf Proof.} We note first that by Lemma \ref{FDElm01} and Remark \ref{FDEnta01} we have
$$
X^{\alpha} \hookrightarrow L^{2}(\mathbb{R}, \mathbb{R}^n)\;\;\mbox{is continuous}.
$$
Now, let $(u_{k}) \in X^{\alpha}$ be a sequence such that $u_{k} \rightharpoonup u$ in $X^{\alpha}$. We will show that $u_{k} \to u$ in $L^{2}(\mathbb{R}, \mathbb{R}^n)$. Suppose, without loss of generality, that $u_{k} \rightharpoonup 0$ in $X^{\alpha}$. The Banach-Steinhaus theorem implies that
$$
A = \sup_{k}\|u_{k}\|_{X^{\alpha}} < +\infty.
$$
For any $y\in \mathbb{R}$, $\forall M>0$ set
\begin{eqnarray*}
I_{M}(y) & = & \{t\in (y-r,y+r)/\;\;l(t) \leq M\},\\
\overline{I}_{M}(y) & = & \{t\in (y-r,y+r)/\;\; l(t) >M\}
\end{eqnarray*}
Choose $\{y_{i}\} \subset \mathbb{R}$ such that $R \subset \cup_{i=1}^{\infty} (y_{i} - r, y_{i} + r)$ and each $t\in \mathbb{R}$ is covered by at most 2 such intervals. Then, for any $M>0$ and $R>2r$, we have
\begin{eqnarray*}
&&\int_{(-R,R)^{c}} |u_{k}(t)|^2dt \leq \sum_{|y_{i}|\geq R-r}^{\infty} \int_{(y_{i}-r, y_{i}+r)} |u_{k}(t)|^2dt\\
&\leq & \sum_{|y_{i}|\geq R-r}^{\infty} \left[ \int_{(y_{i}-r, y_{i}+r) \cap I_{M}(y_{i})} |u_{k}(t)|^2dt + \int_{(y_{i}-r, y_{i}+r)\cap \overline{I}_{M}(y_{i})} |u_{k}(t)|^2dt \right]\\
&\leq &  \sum_{|y_{i}|\geq R-r}^{\infty} \left[ \int_{I_{M}(y_{i})} |u_{k}(t)|^2dt + \frac{1}{M}\int_{(y_{i}-r,y_{i}+r)}l(t)|u_{k}(t)|^2dt \right]\\
&\leq&  \sum_{|y_{i}|\geq R-r}^{\infty} \left[ (\sup_{(y_{i}-r,y_{i} + r)}|u_{k}(t)|)^2 m(I_{M}(y_{i})) + \frac{1}{M} \int_{(y_{i}-r, y_{i} + r)} l(t)|u_{k}(t)|^2dt\right]\\
&\leq&  \sum_{|y_{i}|\geq R-r}^{\infty} \left[ Cm(I_{M}(y_{i})) \|u_{k}\|_{H^{\alpha}(y_{i}-r, y_{i}+r)}^{2} + \frac{1}{M} \int_{(y_{i}-r,y_{i}+r)}l(t)|u_{k}(t)|^2dt\right]\\
&\leq& 2\|u_{k}\|_{X^{\alpha}}^2\left[ C\sup_{|y|\geq R-r}(m(I_{M}(y))) + \frac{1}{M} \right]\\
&\leq & 2 A \left[ C\sup_{|y|\geq R-r}(m(I_{M}(y))) + \frac{1}{M} \right].
\end{eqnarray*}
For any $\epsilon >0$, taking $M$ and $R$ large enough, we can obtain that
$$
\int_{(-R,R)^{c}} |u_{k}(t)|^2dt \leq \frac{\epsilon}{2}.
$$
By Sobolev Theorem, $u_{k} \to 0$ uniformly on $[-R,R]$. Then, for such $R>0$, there exists $k_{0}>0$ such that
$$
\int_{[-R,R]}|u_{k}(t)|^2dt \leq \frac{\epsilon}{2},\:\:\mbox{for all}\;\;k\geq k_{0}.
$$ 
Hence, by the arbitrary of $\epsilon$ we can obtain that $u_{k} \to 0$ in $L^{2}(\mathbb{R}, \mathbb{R}^n)$.
 $\Box$

\begin{Remark}\label{FDEnta02}
From Remark \ref{FDEnta01} and Lemma \ref{FDElm02}, it is easy to verify that the embedding of $X^{\alpha}$ in $L^{q}(\mathbb{R}, \mathbb{R}^n)$ is also continuous and compact for $q\in (2, \infty)$. Therefore, combining this with Theorem \ref{FDEtm01}, for any $q\in [2,\infty]$, there exists $C_{q}$ such that
\begin{equation}\label{FDEeq07}
\|u\|_{L^{q}} \leq C_{q} \|u\|_{X^\alpha}.
\end{equation} 
\end{Remark}

Now, we introduce more notations and some necessary definitions. Let $\mathcal{B}$ be a real Banach space, $I \in C^{1}(\mathcal{B}, \mathbb{R})$, which means that $I$ is a continuously Fr\'echet-differentiable functional defined on $\mathcal{B}$.

\begin{Def}\label{FDEdef01}
$I\in C^{1}(\mathcal{B}, \mathbb{R})$ is said to satisfy the (PS) condition if any sequence $\{u_{j}\}_{j\in \mathbb{N}} \subset \mathcal{B}$, for wich $\{I(u_{j})\}_{j\in \mathbb{N}}$ is bounded and $I'(u_{j}) \to 0$ as $j \to +\infty$, possesses a convergent subsequence in $\mathcal{B}$. 
\end{Def}

In order to find infinitely many solutions of (\ref{Eq02}) under the assumptions of Theorem \ref{tm01}, we shall use the `genus' properties. Therefore, we recall the following definition and result (see \cite{PR}).

Let $\mathcal{B}$ be a Banach space, $I\in C^1(\mathcal{B}, \mathbb{R})$ and $c\in \mathbb{R}$. We set
\begin{eqnarray*}
&&\Sigma = \{A\subset \mathcal{B} \setminus \{0\}:\;\; A\;\;\mbox{is closed in}\;\; \mathcal{B}\;\;\mbox{and symmetric with respect to}\;\; 0\},\\
&& K_{c} = \{u\in \mathcal{B}:\;\; I(u) = c, I'(u) = 0\},\;\;I^{c} = \{u\in \mathcal{B}:\;\; I(u) \leq c\}.
\end{eqnarray*}

\begin{Def}\label{FDEdef02}
For $A\in \Sigma$, we say genus of $A$ is $j$ (denote by $\gamma(A) = j$) if there is an odd map $\psi \in C(A, \mathbb{R}^{j}\setminus \{0\})$, and $j$ is the smallest integer with this property.
\end{Def}

\begin{Lem}\label{FDElm03}
Let $I$ be an even $C^1$ functional on $\mathcal{B}$ and satisfy the (PS) condition. For any $j\in \mathbb{N}$, set
$$
\Sigma_{j} = \{A\in \Sigma:\;\; \gamma (A) \geq j\}, \;\;c_{j} = \inf_{A\in \Sigma_{j}} \sup_{u\in A} I(u).
$$
\begin{enumerate}
\item If $\Sigma_{j} \neq \phi$ and $c_{j} \in \mathbb{R}$, then $c_{j}$ is a critical value of $I$.
\item If there exists $r\in \mathbb{N}$ such that
$$
c_{j} = c_{j+1} = ... = c_{j+r} = c\in \mathbb{R}
$$
and $c\neq I(0)$, then $\gamma (K_{c}) \geq r+1$.
\end{enumerate}
\end{Lem}

\begin{Remark}\label{FDEnta03}
From Remark 7.3 in \cite{PR}, we know that if $K_{c} \in \Sigma$ and $\gamma (K_{c}) >1$, then $K_{c}$ contains infinitely many distinct points, that is, $I$ has infinitely many distinct critical points in $\mathcal{B}$.
\end{Remark}

\section{Proof of Theorem \ref{tm01}}

Now we are in the position to proof Theorem \ref{tm01}. Although its proof is just the repetition of the process of Theorem 1.1 in \cite{ZZRY}, for the reader$'$s convenience, we give some of the details. We begging present some preliminary lemmas.

\begin{Lem}\label{Tlm01}
Under the conditions of $(L_{w})$ and $(HS)_{1}$, if $u_{j} \rightharpoonup u$ in $X^\alpha$, the there exists one subsequence still denoted by $\{u_{j}\}_{j\in \mathbb{N}}$ such that $\nabla W(t,u_{j}) \to \nabla W(t,u)$ in $L^{2}(\mathbb{R}, \mathbb{R}^{n})$.
\end{Lem}

\noindent
{\bf Proof.} Assume that $u_{j} \rightharpoonup u$ in $X^{\alpha}$, then by Banach-Steinhaus Theorem, there exists a constant $M>0$ such that
\begin{equation}\label{Teq01}
\sup_{j\in \mathbb{N}} \|u_{j}\|_{X^{\alpha}} \leq M, \;\;\|u\|_{X^{\alpha}} \leq M.
\end{equation}
Moreover, in view of $(HS)_{1}$, it follows that
\begin{eqnarray}\label{Teq02}
|\nabla W(t, u_{j}(t)) - \nabla W(t,u(t))|^2 &\leq & b^{2}(t) \left(|u_{j}(t)|^{\theta -1} + |u(t)|^{\theta -1} \right)^{2}\\
&\leq& 2b^{2}(t) \left( |u_{j}(t)|^{2(\theta - 1)} + |u(t)|^{2(\theta - 1)} \right)\nonumber
\end{eqnarray}
Therefore, on account of (\ref{FDEeq07}), (\ref{Teq01}), (\ref{Teq02}), and the H\"older inequality, we have
\begin{eqnarray*}
\int_{\mathbb{R}}|\nabla W(t, u_{j}(t)) - \nabla W(t,u(t))|^2dt &\leq& 2\int_{\mathbb{R}}b^{2}(t) \left( |u_{j}(t)|^{2(\theta - 1)} + |u(t)|^{2(\theta - 1)} \right)dt\\
&\leq& 2\|b\|_{L^{\frac{2}{2-\theta}}}^{2} \left( \|u_{j}\|_{L^{2}}^{2(\theta - 1)} + \|u\|_{L^{2}}^{2(\theta - 1)}\right)\\
&\leq& 4C_{2}^{2(\theta -1)} \|b\|_{L^{\frac{2}{\theta - 1}}}^{2} M^{2(\theta -1)}.
\end{eqnarray*}
On the other hand, by Lemma \ref{FDElm02}, $u_{j} \rightharpoonup u$ implies that there exists one subsequence still denoted by $\{u_{j}\}_{j\in \mathbb{N}}$ such that $u_{j} \to u$ in $L^{2}(\mathbb{R}, \mathbb{R}^n)$, which yields that $u_{j}(t) \to u(t)$ for almost every $t\ n \mathbb{R}$. Using Lebesgues's convergence Theorem, we finish the proof of this Lemma. $\Box$

We are going to establish the corresponding variational framework to obtain solutions of (\ref{Eq02}). To this end, define the functional $I:\mathcal{B} = X^{\alpha} \to \mathbb{R}$ by  
\begin{equation}\label{Teq03}
I(u) = \frac{1}{2}\|u\|_{X^\alpha}^{2} - \int_{\mathbb{R}} W(t,u(t))dt. 
 \end{equation}
 
 \begin{Lem}\label{Tlm02}
 Under the conditions of Theorem \ref{tm01}, we have
 $$
 I'(u)v = \int_{\mathbb{R}} \left[ ({_{-\infty}}D_{t}^{\alpha}u(t), {_{-\infty}}D_{t}^{\alpha}v(t)) + (L(t)u(t),v(t))  - (\nabla W(t,u(t)),v(t))\right]dt
 $$
 for all $u,v \in X^{\alpha}$, which yields that
 \begin{equation}\label{Teq04}
 I'(u)u = \|u\|_{X^{\alpha}}^{2} - \int_{\mathbb{R}} (\nabla W(t,u(t)), v(t))dt
 \end{equation}
 \end{Lem}
 
 \noindent
 {\bf Proof.} See \cite{ZZRY}. $\Box$
 
 \begin{Lem}\label{Tlm03}
 If $(L_{w}), (HS)_{1}$ and $(HS)_{2}$ hold, then $I$ satisfies $(PS)$ condition.
 \end{Lem}
 
 \noindent
 {\bf Proof.} Assume that $\{u_{j}\}_{j\in \mathbb{N}} \subset X^{\alpha}$ is a sequence such that $\{I(u_{j})\}_{j\in \mathbb{N}}$ is bounded, and $I'(u_{j}) \to 0$ as $j\to +\infty$. Then there exists a constant $M>0$ such that
 \begin{equation}\label{Teq05}
 |I(u_{j})|\leq M,\;\;\|I'(u_{j})\|_{(X^{\alpha})^{*}} \leq M,
 \end{equation}
 for every $j\in \mathbb{N}$, where $(X^{\alpha})^{*}$ is the dual space of $X^{\alpha}$. Now by (\ref{Teq03}) and (\ref{Teq04}), we obtain that
 \begin{eqnarray}\label{Teq06}
\left(1-\frac{\sigma}{2} \right)\|u_{j}\|_{X^{\alpha}}^{2} &=& I'(u_{j})u_{j} - \sigma I(u_{j})\nonumber\\
&& + \int_{\mathbb{R}}[(\nabla W(t,u_{j}(t)), u_{j}(t)) - \sigma W(t, u_{j}(t))]dt\\
&\leq& M\|u_{j}\|_{X^{\alpha}} + \sigma M.\nonumber
 \end{eqnarray}
 Since $1< \sigma <2$, the inequality (\ref{Teq06}) shows that $\{u_{j}\}_{j\in \mathbb{N}}$ is bounded in $X^{\alpha}$. Then the sequence $\{u_{j}\}_{j\ n \mathbb{N}}$ has a subsequence, again denoted by $\{u_{j}\}_{j\ n \mathbb{N}}$ and there exists $u\in X^{\alpha}$ such that
 $$
 u_{j} \rightharpoonup u\;\;\mbox{in}\;\;X^{\alpha},
 $$
 which yields that
 \begin{equation}\label{Teq07}
 (I'(u_{j}) - I'(u))(u_{j} - u) \to 0.
 \end{equation}
 Moreover, according to Lemma \ref{Tlm01} and the H\"older inequality, we have
 \begin{equation}\label{Teq08}
 \int_{\mathbb{R}} (\nabla W(t,u_{j}(t)) - \nabla W(t,u(t)), u_{j}(t) - u(t))dt \to 0
 \end{equation} 
 as $j\to +\infty$. Consequently, combining (\ref{Teq07}) and (\ref{Teq08}) with the following equality
 $$
 (I'(u_{j}) - I'(u), u_{j} - u) = \|u_{j} - u\|_{X^{\alpha}}^{2} - \int_{\mathbb{R}} (\nabla W(t,u_{j}(t)) - \nabla W(t,u(t)), u_{j}(t) - u(t))dt,
 $$
 it is easy to deduce that $\|u_{j} - u\|_{X^{\alpha}} \to 0$ as $j\to +\infty$. $\Box$
 
 Now, we are in the position to complete the proof of Theorem \ref{tm01}.
 
 \noindent
 {\bf Proof of Theorem \ref{tm01}} According to $(HS)_{1}$ and $(HS)_{3}$, it is obvious that $I$ is even and $I(0) = 0$. In order to apply Lemma \ref{FDElm03}, we prove that
 \begin{equation}\label{Teq09}
 \mbox{for any}\;\;j\in \mathbb{N}\;\;\mbox{there exists}\;\;\epsilon >0\;\;\mbox{such that}\;\; \gamma (I^{-\epsilon}) \geq j.
 \end{equation} 
 
 Let $\{e_{j}\}_{j=1}^{\infty}$ be the standard orthogonal basis of $X^{\alpha}$, that is,
 \begin{equation}\label{Teq10}
 \|e_{j}\|_{X^{\alpha}} = 1\;\;\mbox{and}\;\; \langle e_{i}, e_{k}\rangle_{X^{\alpha}} = 0,\;\;1\leq i \neq k.
 \end{equation}
For any $j\in \mathbb{N}$, define
$$
X_{j}^{\alpha} = \mbox{span}\{e_{1}, e_{2},...,e_{j}\},\;\;S_{j} = \{u\in X_{j}^{\alpha}:\;\;\|u\|_{X^{\alpha}} = 1\},
$$ 
then, for any $u\in X_{j}^{\alpha}$, there exists $\lambda_{i} \in \mathbb{R}$, $i = 1,2,...,j$, such that
\begin{equation}\label{Teq11}
u(t) = \sum_{i=1}^{j}\lambda_{i}e_{i}(t)\;\;\mbox{for}\;\;t\in \mathbb{R},
\end{equation}
which indicates that
\begin{equation}\label{Teq12}
\|u\|_{L^{\theta}} = \left(\int_{\mathbb{R}} |u(t)|^{\theta} \right)^{1/\theta} = \left( \int_{\mathbb{R}} |\sum_{i=1}^{j} \lambda_{i}e_{i}(t)|^{\theta}dt \right)^{1/\theta}
\end{equation} 
and 
\begin{eqnarray}\label{Teq12-1}
\|u\|_{X^{\alpha}}^{2} &=&\int_{\mathbb{R}} \left[ |{_{-\infty}}D_{t}^{\alpha}u(t)|^2 + (L(t)u(t), u(t)) \right]dt\nonumber\\
&=& \sum_{i=1}^{j} \lambda_{i}^{2} \int_{\mathbb{R}} \left[|{_{-\infty}}D_{t}^{\alpha}e_{i}(t)|^2 + (L(t)e_{i}(t), e_{i}(t))\right]dt\\
& = & \sum_{i=1}^{j} \lambda_{i}^{2} \|e_{i}\|_{X^{\alpha}}^{2} = \sum_{i=1}^{j} \lambda_{i}^{2}\nonumber.
\end{eqnarray}

On the other hand, in view of $(HS)_{1}$, for any bounded open set $D\subset \mathbb{R}$, there exists $\eta >0$ (dependent on $D$) such that
\begin{equation}\label{Teq13}
W(t,u) \geq a(t)|u|^{\theta} \geq \eta |u|^{\theta},\;\;(t,u) \in D\times \mathbb{R}^{n}.
\end{equation}
As a result, for any $u\in S_{j}$, we can take some $D_{0} \subset \mathbb{R}$ such that
\begin{equation}\label{Teq14}
\int_{\mathbb{R}} W(t, u(t))dt = \int_{\mathbb{R}} W(t, \sum_{i=1}^{j}\lambda_{i}e_{i}(t))dt \geq \eta \int_{D_{0}} |\sum_{i=1}^{j} \lambda_{i}e_{i}(t)|^{\theta}dt = \varrho >0.
\end{equation} 
Indeed, if not, for any bounded open set $D\subset \mathbb{R}$, there exists $\{u_{n}\}_{n\in \mathbb{N}} \in S_{j}$ such that
$$
\int_{D}|u_{n}(t)|^{\theta}dt = \int_{D} \left|\sum_{i=1}^{j} \lambda_{in} e_{i}(t)\right|^{\theta}dt \to 0,\;\;\mbox{as}\;\;n\to 0,
$$ 
where $u_{n} = \sum_{i=1}^{j} \lambda_{in} e_{i}$ such that $\sum_{i=1}^{j} \lambda_{in}^{2} = 1$. Because $\sum_{i=1}^{j} \lambda_{in}^{2} = 1$, we have
$$
\lim_{n\to +\infty} \lambda_{in} =: \lambda_{i0} \in [0,1]\;\;\mbox{and}\;\; \sum_{i=1}^{j}\lambda_{i0}^{2} = 1.
$$ 
Hence, for any bounded open set $D\subset \mathbb{R}$,
$$
\int_{D} \left| \sum_{i=1}^{j} \lambda_{i0}e_{i}(t) \right|^{\theta}dt = 0.
$$ 
The fact that $D$ is arbitrary yields that $u_{0} = \sum_{i=1}^{j} \lambda_{i0}e_{i}(t) = 0$ a.e. on $\mathbb{R}$ which contradicts the fact that $\|u_{0}\|_{X^{\alpha}} = 1$. Hence, (\ref{Teq14}) holds true.

In addition, because all norms of a finite dimensional norm space are equivalent, there is a constant $c' >0$ such that
\begin{equation}\label{Teq15}
c'\|u\|_{X^{\alpha}} \leq \|u\|_{L^{\theta}},\;\;\forall u\in X_{j}^{\alpha}.
\end{equation} 
Consequently, according to $(HS)_{1}$ and (\ref{Teq11})-(\ref{Teq15}), we have
\begin{eqnarray*}
I(su) & = & \frac{s^2}{2}\|u\|_{X^\alpha}^{2} - \int_{\mathbb{R}} W(t, su(t))dt\\
& = & \frac{s^2}{2}\|u\|_{X^\alpha}^{2} - \int_{\mathbb{R}} W(t, s\sum_{i=1}^{j} \lambda_{i} e_{i}(t))dt\\
&\leq& \frac{s^2}{2}\|u\|_{X^\alpha}^{2} - s^{\theta} \int_{\mathbb{R}} a(t) \left| \sum_{i=1}^{j} \lambda_{i} e_{i}(t)\right|^{\theta}dt\\
&\leq & \frac{s^2}{2}\|u\|_{X^\alpha}^{2} - \eta s^{\theta} \int_{D_{0}} \left| \sum_{i=1}^{j} \lambda_{i} e_{i}(t) \right|^{\theta}dt  \\
&\leq & \frac{s^2}{2}\|u\|_{X^\alpha}^{2} - \varrho s^{\theta}\\
& = & \frac{s^2}{2} - \varrho s^{\theta},\;\; u \in S_{j},
\end{eqnarray*}
which implies that there exists $\epsilon >0$ and $\delta >0$ such that
\begin{equation}\label{Teq16}
I(\delta u) < -\epsilon\;\;\mbox{for}\;\;u\in S_{j}.
\end{equation}
Let
$$
S_{j}^{\delta} = \{\delta u/\;\;u\in S_{j}\},\;\;\Omega = \left\{(\lambda_{1}, \lambda_{2},...,\lambda_{j})  :\;\;\sum_{i=1}^{j}\lambda_{i}^{2} < \delta^{2}\right\}.
$$
The it follows from (\ref{Teq16}) that
$$
I(u) <-\epsilon, \;\;\forall u\in S_{j}^{\delta},
$$
which, together with the fact that $I\in C^{1}(X^{\alpha}, \mathbb{R})$ and is even, yields that
$$
S_{j}^{\delta} \subset I^{-\epsilon} \in \Sigma.
$$
On the other hand, it follows from (\ref{Teq11}) and (\ref{Teq12-1}) that there exists an odd homeomorphism mapping $\varphi \in C(S_{j}^{\delta}, \partial \Omega)$. By some properties of the genus, we obtain
\begin{equation}\label{Teq17}
\gamma (I^{-\epsilon}) \geq  \gamma (S_{j}^{\delta}) = j,
\end{equation}
so (\ref{Teq09}) follows. Set
$$
c_{j} = \inf_{A\in \Sigma_{j}} \sup_{u\in A} I(u),
$$
then, from (\ref{Teq17}) and the fact that $I$ is bounded from below on $X^{\alpha}$, we have 
$$-\infty < c_{j} \leq -\epsilon <0,$$
that is, for any $j\in \mathbb{N}$, $c_{j}$ is a real negative number. By Lemma \ref{FDElm03} and Remark \ref{FDEnta03}, $I$ has infinitely many nontrivial critical points, and consequently, (\ref{Eq02}) possesses infinitely many solutions.

In what follows, we show that $c_{j} \to 0^{-}$ as $j\to +\infty$. To this end, define
$$
X_{j} = span \{e_{j}\}, \;\;Z_{j} = \overline{\bigoplus_{k=j}^{\infty} X_{k}},
$$
where $\{e_{j}\}_{j=1}^{\infty}$ is the standard orthogonal basis of $X^{\alpha}$. Set
\begin{equation}\label{Teq18}
\beta_{j} = \sup_{u\in Z_{j}, \|u\|_{X^{\alpha}}=1}\|u\|_{L^2},
\end{equation} 
then $\beta_{j} \to 0$ as $j\to +\infty$. Indeed, it is clear that $0< \beta_{j+1} \leq \beta_{j}$, so that $\beta_{j} \to \beta \geq 0$ as $j\to +\infty$. For every $j\geq 1$, there exists $u_{j}\in Z_{j}$ such that $\|u_{j}\|_{X^{\alpha}} = 1$ and $\|u_{j}\|_{L^2} \geq \frac{\beta_{j}}{2}$. By definition of $Z_{j}$, $u_{j} \rightharpoonup 0$ in $X^{\alpha}$. Then it implies that $u_{j} \to 0$ in $L^{2}(\mathbb{R}, \mathbb{R}^{n})$ by Lemma \ref{FDElm02}. Thus, we have achieved that $\beta =0$. In view of (\ref{Eq04}), (\ref{FDEeq07}), and the H\"older inequality, we have
\begin{eqnarray}\label{Teq19}
0\leq \int_{\mathbb{R}} W(t,u(t))dt & \leq & \frac{1}{\theta}\int_{\mathbb{R}} b(t)|u(t)|^{\theta}dt\nonumber\\
&\leq& \frac{1}{\theta} \|b\|_{L^{\frac{2}{2-\theta}}} \|u\|_{L^{2}}^{\theta}\nonumber \\
&\leq& \frac{C_{2}^{\theta}}{\theta} \|b\|_{L^{\frac{2}{2-\theta}}} \|u\|_{X^{\alpha}}^{\theta}.
\end{eqnarray}
Therefore by (\ref{Teq03}) and (\ref{Teq19}), we obtain
$$
I(u) \geq \frac{1}{2} \|u\|_{X^\alpha}^{2} - \frac{C_{2}^{\theta}}{\theta} \|b\|_{L^{\frac{2}{2-\theta}}} \|u\|_{X^\alpha}^{\theta},
$$
which yields that $I(u)$ is coercive, that is, $I(u) \to +\infty$ as $\|u\|_{X^{\alpha}} \to +\infty$. Therefore, there exists a $\tau >0$ such that $I(u) >0$ for $\|u\|_{}X^{\alpha} \geq \tau$. Moreover, for any $A\in \Sigma_{j}$, because $\gamma (A) \geq j$, we have $A\cap Z_{j} \neq \phi$. Subsequently, on account of (\ref{Teq18}), it indicates that
\begin{eqnarray*}
\sup_{u\in A} I(u) \geq \inf_{u\in Z_{j},\;\|u\|_{X^{\alpha}}\leq \tau} I(u)&\geq& \inf_{u\in Z_{j},\;\|u\|_{X^{\alpha}}\leq \tau} \left( \frac{1}{2}\|u\|_{X^\alpha}^{2} - \frac{\beta_{j}^{\theta}}{\theta}\|b\|_{L^{\frac{2}{2-\theta}}}\|u\|_{X^{\alpha}}^{\theta} \right)\\
&\geq& - \frac{\beta_{j}^{\theta}}{\theta} \|b\|_{L^{\frac{2}{2-\theta}}}\tau^{\theta},
\end{eqnarray*}
which implies that
$$
c_{j} = \inf_{A\in \Sigma_{j}} \sup_{u\in A}I(u) \geq -\frac{\beta_{j}^{\theta}}{\theta} \|b\|_{L^{\frac{2}{2-\theta}}}\tau^{\theta}.
$$
Combining this with $c_{j}>0$ and $\beta_{j} \to 0$, we get $c_{j} \to 0^{-}$ as $j\to +\infty$. $\Box$


\noindent {\bf Acknowledgements:}
This work was supported by myself.

\end{document}